\documentclass[journal]{IEEEtran}
\usepackage{amsmath,graphicx,mathtools,amsthm}
\usepackage[utf8]{inputenc}
 \usepackage[switch,pagewise]{lineno}
\usepackage{epstopdf}
\usepackage{algorithm}
\usepackage{algorithmic}
\usepackage{amssymb}
\usepackage{extarrows}
\usepackage{graphicx}
\makeatletter
\def\fps@eqnfloat{t}
\def\ftype@eqnfloat{4}

\newenvironment{eqnfloat*}
{\@dblfloat{eqnfloat}} {\end@dblfloat} \makeatother
\usepackage{multirow}
\usepackage{booktabs}
\usepackage{pgfplots}
\usepackage{tikz}
\usepgflibrary{arrows.meta}

\graphicspath{{Figures/}}
\usepackage{float}
\usepackage{cite}
\usepackage{epsfig,latexsym}
\usepackage{breqn}
\usepackage{subcaption}
\usepackage{lipsum}
\usepackage{xcolor}
\usepackage{bm}
\usepackage[british]{babel}
\usepackage{color, colortbl} 
\definecolor{Gray}{gray}{0.9} 

\usepackage{epstopdf}

\def\MSK{\textrm{MSK}}
\def\DPSK{\textrm{DPSK}}
\def\PRI{\textrm{PRI}}

\usepackage{bm}
\newcommand{\xdownarrow}[2][]{%
\left.{#1}\right\downarrow{#2}}
\usepackage{blindtext, rotating}
\usepackage{csquotes}
\newcommand{\vectonorm}[1]{\left|\left|#1\right|\right|}

\usepackage{mathtools,amsthm}
\newtheoremstyle{case}{}{}{}{}{}{:}{ }{}
\theoremstyle{case}

\usepackage{cite}
\usepackage{epsfig,latexsym}
\usepackage{amsmath}

\usepackage{soul}

\title{Generalized Fully Coherent Closed-form Receiver Design for Joint Radar and Communication System}
\author{Muhammad Zubair, Sajid Ahmed, {\em Senior Member, IEEE},  and Mohamed-Slim Alouini, {\em Fellow, IEEE}\\
King Abdullah University of Science and Technology\\
zubair.chattha79@gmail.com, sajid.ahmed@kaust.edu.sa, slim.alouini@kaust.edu.sa.
}

\begin{document}
%
\maketitle
\begin{abstract}
In conventional radar, the transmission of the same waveform is repeated after a predefined interval of time called pulse-repetition-interval (PRI). This technique helps to estimate the range and Doppler shift of targets and suppress clutter. 
In dual-function radar communication (DFRC), different waveforms are transmitted after each PRI. Thus, each waveform yields different range-side-lobe (RSL) levels at the receiver's output. As a consequence, Doppler shift estimation and clutter suppression become challenging tasks. A state-of-the-art (SOTA) method claims that if the number of waveforms is more than two, it is impossible to achieve fully coherent RSL levels with both waveforms. Therefore, this algorithm uses iterative methods to achieve as much as possible coherency and minimize the RSL levels. In contrast to that SOTA method, we proposed two novel closed-form receivers for the DFRC that yield a fully coherent response for several waveforms and suppress the RSL levels. Experimental results demonstrate that the proposed receivers achieve full coherency and the RSL levels are significantly lower than the conventional method. 
\end{abstract}  

\section{Introduction}\label{sec:Intro}
{\huge I}n wireless communication, the radio spectrum is the most precious and expensive resource. Radio spectrum extending from 1MHz to 300GHz has favorable propagation characteristics, therefore, this spectrum is being used for different purposes. The use of this band for different applications is summarized in Table \ref{tab:RadarSpectrum}. There are currently 7.9 billion mobile subscriptions and it is predicted that the total mobile subscriptions by the end of 2026 will reach 8.8 billion. Similarly, the overall Internet-of-Things (IoT) connections have reached 12.6 billion and in 2026 they are expected to reach 26.9 billion. The fifth-generation (5G) standard was introduced in 2020 to accommodate the high demand for subscriptions within the available spectrum. The 5G will be much faster, reliable, more efficient, and will support data rates up to 35.46 Gbps \cite{6G_SlimAlouni2020}. It will use frequency range 1 (FR1), also known as a sub-6GHz band (2-7.125GHz), and frequency range 2 (FR2), known as mm-Wave radio band (24.25-52.6GHz), as shown in Table. \ref{tab:RadarSpectrum}. 
Future applications, such as holographic calls and flying networks, will require even more high data rates. In this perspective quest for 6G and 7G has been already started \cite{6G_SlimAlouni2020,6G_zong2019,6G_david2018,6G7G_gawas2015,6G7G_mondal2015}. In the 6G and 7G,  the integration of 5G, satellite network, and space roaming will be carried out. Spacecraft, drones, and radars will be considered a simple IoT devices \cite{Spacecraft_as_Iot_2018searching,Spacecraft_as_IoT,Drones_as_IoT_akan2020internet}. 

As cellular communication systems have evolved from 1G to 5G (please see Table \ref{tab:RadarSpectrum}) and on the way to 6G, radar systems are also evolving at a much faster pace. Until the second world war, radar systems had a significant size, were bulky, power-hungry, and the cost was enormous. Moreover, the use of radar was restricted to the military only. Now, the size and cost of radar have significantly reduced, and it is entering the consumer market. For example, now the radars are being used in the automotive industry, medical devices for non-invasive testing, gesture control smart devices, and wearable devices  \cite{MedicalRadar1,MedicalRadar2,MedicalRadar3,soliRadar_Google2016}.

Until now, radar and cellular communication systems have been developed and studied separately. Both systems use their frequency spectrum . In Table \ref{tab:RadarSpectrum}, it is summarized how radar bands for different applications underutilized the limited radio spectrum. Now the evolution of applications requiring high data rate communication has completely occupied the limited radio spectrum and started to overlap with existing radar bands as shown in Table \ref{tab:RadarSpectrum}. The scarcity of radio spectrum urged researchers to develop new solutions to overcome this problem. One of the possible solutions is the integration of both systems. As both technologies rely on the same electromagnetic radiation phenomena, the same equipment, spectrum, and signals can be used for both systems. For example, suppose that the vicinity objects reflect the transmitted signal to convey information from point A to B. These reflected signals from objects can be received at point A and processed to estimate the position and velocity of various objects. Therefore, to efficiently use the available spectrum for both communication and radar, extensive research is being done to devise new techniques. The spectrum sharing and the use of identical waveforms for communications and radar will likely be the main feature of future communications systems in 5G and beyond. 
\begin{table*}[]
    \centering
\begin{tabular}{ |p{1.6cm}|p{2cm}|p{1.8cm}|p{5cm}|p{2.5cm}|  }
 \hline 
 \rowcolor{Gray}
\textbf{Band}  &    \textbf{Frequencies}  & \textbf{Wavelengths} & \textbf{Applications}  & \textbf{Evolution; 1G-5G and beyond}\\ 
\hline
HF            &    3 - 30MHz             & 100 - 10m  &  Over the horizon surveillance radar       &                \\
\cline{1-4} 
VHF           &    30 - 300MHz           & 10 - 1m & Long range surveillance, Counter stealth  &     {\color{blue} \multirow{4}{*}{
$\xdownarrow[\begin{gathered} 
\hfill
\rotatebox{90}{4G, 3G, 2G, 1G }
\end{gathered}]{}$ 
 }
 }
\multirow{4}{*}{$\xdownarrow[\begin{gathered}
\hfill \\
\rotatebox{90}{5G(7.125GHz )}\\
\end{gathered}]{\rotatebox{90}{FR1}}$ }            \\
\cline{1-4} 
UHF           &    300MHz - 1GHz         & 1 - 30cm           &   Long range surveillance, Counter stealth    &                  \\
\cline{1-4} 
L            &    1 - 2GHz               & 30 - 15cm          &    Air traffic control   &                   \\
\cline{1-4} 
S            &    2 - 4GHz               & 15 - 7.5cm         &  Weather observation     &                   \\
\cline{1-4} 
C            &    4 - 8GHz               &7.5 - 3.75cm        &  Short range, Missile guidance     &                   \\
\hline
X            &    8 - 12GHz              &3.75 - 2.5cm        &   Unmanned aerial vehicle, \textbf{Mobile/Fixed service on primary basis}    &      {\color{red}FR3:} More challenges due to attenuation, currently under study by ITU.   \\
\cline{1-4} 
K\textsubscript{u}&    12 - 18GHz       & 2.5 - 1.67cm        &  Police radar, \textbf{Mobile/Fixed service on primary basis}       &                   \\
 \hline 
K           &    18 - 27Ghz             & 1.67 - 1.11cm       &    Airport surveillance   &           
    \multirow{4}{*}{$\qquad \quad \xdownarrow[\begin{gathered}
  \rotatebox{90}{5G(mm Wave)}
   \\
  \end{gathered}]{\rotatebox{90}{FR2}}$ }\\
 \cline{1-4} 
K$_a$&    27 - 40 Ghz       &1.11 - 7.5mm        & Scientific remote sensing      &                  \\ 
 \cline{1-4} 
mm           &    40 - 52.6GHz           & 7.5 - 5.7mm          &  Automobile cruise control,   \textbf{NR-U, NR-Light}   &                  \\
  \cline{1-4}
{FR2+}\textsuperscript{4)}           &   57 - 71GHz          & 5.3 - 4.2mm          &  Automobile cruise control, \textbf{NR-Light}    &                  \\
\hline
FR4           &   52.6 - 114.25GHz            & 5.7 - 2.6mm          &  Automobile cruise control     &     \\
\hline
 FR5 (D band)           &    110 - 170 GHz           & 2.7 - 1.8mm          &  Automobile cruise control,  \textbf{Fundamental 6G research }     &    \textcolor{red}{\multirow{4}{*}{$\qquad \quad \xdownarrow[ \begin{gathered}
  \rotatebox{90}{6G(sub-THz)}
   \\
  \end{gathered}]{\rotatebox{90}{FR5}}$ }}\\[5ex]
 FR5 (G band)           &    140 - 220 GHz           & 2.1 - 1.4mm          &  Automobile cruise control,  \textbf{Fundamental 6G research}     &\\
 \hline
  FR5             &    220 - 275GHz           & 2.1 - 1.1mm          &  Automobile cruise control     & \\
 \hline
          &    252 - 325GHz           & 1.2 - 0.923mm          &  Automobile cruise control,  \textbf{ 802.15.3d-2017 (P2P communication with 100 Gbps)}     & \textcolor{red}{\multirow{2}{*}{$\qquad \quad  \xdownarrow[\begin{gathered} 
  \rotatebox{90}{6G (THz) }\hfill \\
  \end{gathered}]{}$ }} \\ [4ex]
              &    0.3 - 3THz           &  1mm - 100um          &  Automobile cruise control     & \\
 \hline
\end{tabular}
    \caption{Frequency spectrum overlapping by radar and communication. In application column bold and normal fonts respectively represent use of frequency band for communication and radar \cite{Spectrum_bruder2003ieee,docomo2020white}.  {\textbf {FR2+}\textsuperscript{4)} represents the addition of FR4 frequency band from 52.6GHz to 71GHz to FR2 for the evolution of 5G applications named as New radio Light(NR-Light) or reduced capability (RedCap) devices. 
    }}
    \label{tab:RadarSpectrum}
\end{table*}

In the literature, there are mainly four categories of joint radar and communication research, which are as follows:
\begin{enumerate}
   
\item \textbf{Spectrum Sharing:} 
In this technique, both communication and radar systems share the same spectrum when it is free from the other \cite{spectrumSharing1_cabric2006,spectrumSharing2_deng2013,spectrumSharing3_Khawar2012,spectrumSharing4_Himed2013,SpectrumSharing5_2018jointDesign,SpectrumSharing6_2018mimoliu}. 

\item \textbf{Communication Centric Joint Communication Radar (JCR):} 
In JCR, radar parameters are estimated from the transmission of the communication system \cite{JCR4,JCR1,JCR2,JCR3}. 

\item \textbf{Radar Centric Joint Radar Communication (JRC):} 
In JRC technique, communication is realized on a primary radar system. The pioneering work in this respect is presented in \cite{JRC1_MIMORadarcommOFDMPioneer2006,JRC3_arik2019}, where OFDM communication is employed on MIMO radar. 

\item \textbf{dual function radar communication (DFRC):} 
In DFRC, radar and communication functions are implemented on a single hardware platform. Here, a single waveform serve the purpose of both data communication and radar parameter estimation. Based on the system model, DFRC can be further classified into two categories: i) Communication Centric DFRC \cite{CC_DFRC1_liu__Waveform2018,CC_DFRC2_liu__reducedSideLobes_Waveform2020,CC_DFRC3_liu_stateOfArt_roadAhead2020,CC_DFRC4_liu2020tutorial} , and ii) Radar Centric DFRC \cite{CC_DFRC5_ni2020parameter,CC_DFRC6_luo2019optimization,CC_DFRC7_kumari2017ieee,RC_DFRC2_sturm2011waveform,DFRC_Rx_2010,RC_DFRC3_hassanien2015,wang2018MIMOdual,RC_DFRC4_hassanien2019,RC_DFRC5ahmed2019distributed,RC_DFRC6_ammar2019ofdm,RC_DFRC8_ammar2018multi}. Both of these technologies rely either on the design of receive filter or waveform design. 
\end{enumerate}

Although, all of the above techniques, i.e., spectrum sharing, JCR, and JRC have their significance but DFRC due to to its single hardware for dual functionality has more applications compared to the other methods. Therefore, the focus of this paper will be on DFRC, and its unaddressed problem will be solved. 

For example, the biggest problem that arises in the DFRC system is the detection and estimation of radar parameters.  Because, due to vast radar-cross-section (RCS) and spread of clutter (the electromagnetic return from mountains, ground, and buildings ), it shows high peak values in the range-side-lobes(RSLs) that can mask weak targets. Moreover, due to different waveforms in CPI, RSLs are not coherent that can degrade Doppler shift estimation.  To address this problem, state-of-the-art (SOTA) research designs unique waveforms or receive filters for the suppression of the RSLs, and coherent response for different waveforms.
For instance, to design waveforms for DFRC, one closed-form and two iterative algorithms are proposed in \cite{CC_DFRC1_liu__Waveform2018}. The proposed algorithms satisfy constant modulus and similarity constraints. However, none of them addresses the suppression of RSLs. Therefore, in \cite{CC_DFRC2_liu__reducedSideLobes_Waveform2020} to reduce RSLs, integrated-side-lobes constraint is incorporated in the objective function. Similarly, further developments on DFRC waveforms are presented in \cite{RC_DFRC3_hassanien2015,wang2018MIMOdual,RC_DFRC4_hassanien2019,RC_DFRC5ahmed2019distributed,RC_DFRC6_ammar2019ofdm,RC_DFRC8_ammar2018multi,CC_DFRC1_liu__Waveform2018,CC_DFRC2_liu__reducedSideLobes_Waveform2020,CC_DFRC3_liu_stateOfArt_roadAhead2020,CC_DFRC4_liu2020tutorial,CC_DFRC5_ni2020parameter,CC_DFRC6_luo2019optimization,CC_DFRC7_kumari2017ieee,RC_DFRC2_sturm2011waveform,RC_DFRC_shannonWaveforms}.
For DFRC, iterative receive filters are proposed in \cite{RC_DFRC_shannon_Waveforms1,RC_DFRC_shannon_Waveforms2,RC_DFRC_shannon_Waveforms3}. The proposed iterative receive filters suppress the RSLs, but the RSLs of receive filters corresponding to different waveforms are non-coherent . Due to this problem, they fail to perform well.  
\\
\\
{\bf Contributions:}
Two main contribution of this paper are:
\begin{enumerate}
\item In this work, we transformed the problem into a constrained optimization problem that guarantees a fully coherent receiver for any number of waveforms. While the existing research claims that the design of the fully coherent receiver is not possible for more than two waveforms.  

\item Two generalized closed-form receivers are designed based on linear and circular convolution model for multiple waveforms. The output response of both receivers is fully coherent and the RSL levels are also lower than the existing iterative method.  
\end{enumerate}
{\bf Paper Organization:}
\\
The rest of the paper is organized as follows: The  problem formulation is discussed in Sec. \ref{Sec:SigMod}, proposed closed-form receivers design is described in Sec. \ref{Sec:Proposed}. The receive signal model for radar and communication is discussed in Sec. \ref{Sec:ClutterModel}. Experimental results are discussed in Sec. \ref{Sec:Simulations}, and conclusions are drawn in Sec. \ref{Sec:Simulations}.
\\
\textbf{Notations:} Bold upper case letters, ${\bf X}$ represents matrices and lower case letters, ${\bf x}$, denote vectors. Transpose and conjugate transposition of a matrix are denoted by $(\cdot)^T$ and $(\cdot)^H$, respectively. The conjugate of a scalar is denoted by $(\cdot)^*$ . Convolution operator is denoted by $*$ and the close interval $\{x: a \le x \le b\}$ is denoted as $[a,b]$.  

\section{Problem Formulation} \label{Sec:SigMod}
\begin{figure*}[t]
    \centering
    \includegraphics[width=7in,keepaspectratio]{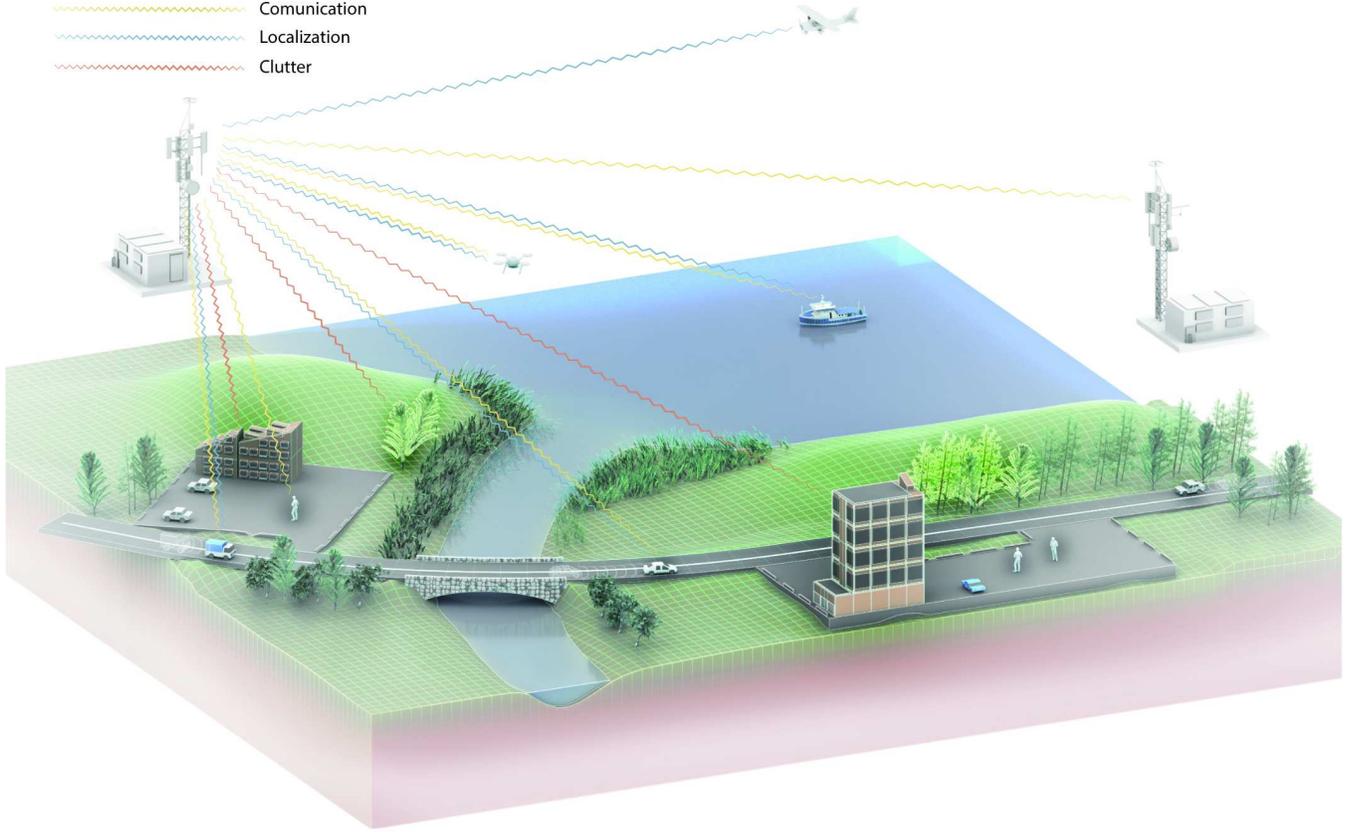}
    \centering
    \caption{Joint Radar-Communication scenario. Transmitted signal from the base-station is used for radar (blue waveform) and communication (yellow waveform) functions. The reflected signal from the stationary objects is considered as a clutter signal.\\
    Authors are thankful for the illustration created by Heno Hwang. Scientific Illustrator at Research Communication and Publication Services. Office of the Vice President for Research - King Abdullah University of Science and Technology.}  
    \label{fig:KAUST}
\end{figure*}

\begin{figure}
    \centering
    \includegraphics{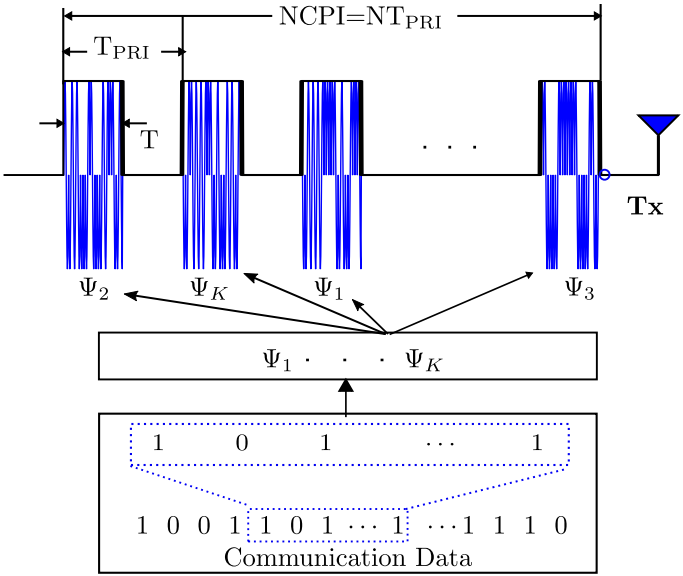}
    \caption{Embedding data bits in radar pulses. Here, $\Psi_k$ is the $k${th} waveform, $T$ is the pulse duration, $T_{\PRI}$ is pulse repetition interval, and NCPI is the non-coherent processing interval.}
    \label{fig:problem_formulation}
\end{figure}

Consider a joint radar-communication scenario shown in Fig. \ref{fig:KAUST}, where a base station (BS) intends to communicate with multiple mobile-users (MU) and localize various targets using the transmitted waveforms for communication. Depending on the antenna's beampattern, the transmitted signal propagates in different directions. It is received by multiple objects, such as mobile user, vehicle, aeroplane, ship, ground, and buildings, present at different range bins. In the figure \ref{fig:KAUST}, blue and yellow waveforms respectively represent the transmitted signal parts for radar and communication, while the red waveform represents the transmitted signal received by the stationary objects. If the RCS of an object is sufficiently large, part of the transmitted signal energy will be reflected back and it can be used to localize that object. A significant portion of the reflected signal energy is from the ground and buildings due to their large RCS. However, we are not interested in localizing them. Therefore, the received signal from such targets is called clutter. Due to the vast RCS of clutter compared to the objects we are interested in, it becomes challenging to estimate these object's parameters. However, the exploitation of Doppler shift of moving objects can help to localize them.
In conventional radar systems, a coherent-processing-interval (CPI) technique is used to estimate the target's Doppler shift. In this technique, $N$ identical waveforms are transmitted in each CPI. Receiver's output identical index samples from $N$ waveforms are called slow time samples. To estimate the Doppler shift of a target fast-Fourier-transform (FFT) is applied on slow time samples in the CPI~\cite{Radarfundamentalsrichards2014}. However, for the joint radar-communication system, a non-coherent processing interval (NCPI) technique is required that can allow $N$ different waveforms in each processing interval to transmit various communication bits in each waveform. For the transmission of $M$ symbols, the NCPI technique is illustrated in Fig. \ref{fig:problem_formulation}. For a particular symbol, the system can select a corresponding waveform to transmit from $K$ waveforms. Each waveform/symbol carries $\log_2(K)$ bits. In the NCPI technique, since waveforms are different, it becomes challenging to estimate Doppler shift by applying FFT on slow time samples. The second issue arises due to the range side-lobes. After matched-filtering, clutter due to its high RCS can produce high range side-lobe values that can mask small targets. Moreover, different waveforms can have different range side-lobe values for the same clutter, affecting the receiver's sensitivity, especially for small targets. Therefore, for each waveform, range side-lobe values should be coherent and as-small-as possible. To address this problem, in the sequel, we design receive filters for $K$ waveforms and constrain them to yield coherent output responses. 

If $\Psi_k(t)$ is the transmitted waveform, $h_k(t)$ is the corresponding receive filter to be designed, and $e_k(t)$ is the desired output response, mathematically, the receive filter design problem can be expressed as
\begin{eqnarray}
\Psi_{k}(t)*h_{k}(t) = e_k(t), \quad k = 1,2,\ldots,K \hspace{5em} \notag\\
\mbox{subject to constraint} \hspace{13em} \notag\\
\Psi_{1}(t)*h_{1}(t)=\Psi_{2}(t)*h_{2}(t)=\cdots=\Psi_{K}(t)*h_{K}(t).
\label{eq:COP}
\end{eqnarray}
The solution of (\ref{eq:COP}) yields optimal receive filters that can produce coherent response for non-coherent waveforms. 
The performance of the filter is assessed for differential phase shift keying (DPSK) and minimum-phase-shift keying (MSK) symbols. Due to continuous phase change, DPSK and MSK waveform gives narrow bandwidth and good auto-correlation compared to the other phase modulation schemes.
DPSK waveforms can be designed in two steps. In the first step, a digital constant envelope waveform using binary random sequence is generated as

\begin{eqnarray}
s_d(t) = \sum_{n=0}^{N-1}e^{j\pi\left(\frac{x(n)+1}{2}\right)}\left[u\left(t-n\tau_{c}\right) -u\left(t-(n+1)\tau_{c}\right)\right] \notag, 
\end{eqnarray}
where $x(n)\in \pm 1$ generated from Gaussian distribution, $u(t)$ is the unit step function, and $\tau_{c}$ is chip interval. While in the second step, pulse shaping is applied across bit boundaries to convert discontinuous phase transition to continuous. Thus final DPSK waveform for baseband frequency $f_b =\frac{1}{2\tau_{c}}$ can be expressed as
\begin{align}
 \Psi_{\DPSK}(t) &= s_d\left(t-\frac{\tau_{c}}{2}\right)\left|\cos \left(2\pi f_bt \right)\right|+ js_d(t)\left|\sin\left(2\pi f_bt  \right)\right|\notag\\
     &= s_c\left(t-\frac{\tau_{c}}{2}\right) +j s_s(t) \quad, 0\leq t\leq T
    \label{eq:DPSK},
\end{align}
where $s_{c}(t) = s_d\left(t-\frac{\tau_c}{2}\right)\left|\cos(2\pi f_bt)\right|$ and $s_{s}(t) = s_d(t)\left|\sin(2\pi f_bt)\right|$. 
If $\tau_c$ is a single chip interval, for $N$ chips, $T = N\tau_{c}$. 
The time domain passband version of complex envelope or baseband signal defined in (\ref{eq:DPSK}) is as follows,
\begin{align}
    \Psi_p(t) &= s_{\mbox{\tiny{c}}}\left(t-\frac{\tau_{c}}{2}\right) \cos\left(2\pi f_{c}t\right) - {s}_{\mbox{\tiny{s}}}(t)\sin(2\pi f_{c}t ).
    \label{eq:pasbnd1}
\end{align}
Now the relationship between passband and baseband in exponential form can be expressed as,
\begin{align}
  \Psi_p(t) &= \mbox{Re}\left(\Psi_{\mbox{\tiny{DPSK}}}(t) e^{j2\pi f_{c}t}\right)
  \label{eq:pasbnd2}
\end{align}
%
\textbf{Envelope and Phase of Passband Signal:} 
\\
The baseband signal $\Psi_p(t)$ can be expressed in polar form. Define the envelope and phase $\theta(t)$ as
\begin{align}
    \left|\Psi_{\mbox{\tiny{DPSK}}}(t)\right|&=  \sqrt{ {s}_{\mbox{\tiny{c}}}^{2}\left(t-\frac{\tau_{c}}{2}\right) + {s}_{\mbox{\tiny{s}}}^{2}(t)) },\notag\\
    \theta & =\tan^{-1}\left( \frac{{s}_{\mbox{\tiny{s}}}(t)}{{s}_{\mbox{\tiny{c}}}(t-\frac{\tau_{c}}{2})}\right)\notag.
\end{align}
Substituting $\Psi_{\mbox{\tiny{DPSK}}}(t)=\left|\Psi_{\mbox{\tiny{DPSK}}}(t)\right|e^{j\theta(t)}$ into (\ref{eq:pasbnd2}), we get another form of passband signal
\begin{align}
    \Psi_{\mbox{\tiny{P}}}(t) = \left|\Psi_{\mbox{\tiny{DPSK}}}(t)\right|\cos(2\pi f_{c}t + \theta(t)).
    \label{eq:pasbnd3}
\end{align}
Similar to DPSK, minimum-shift-keying (MSK) can also be used. MSK uses binary codes and can be expressed as \cite{DFRC_Rx_2010} 
\begin{equation}
     \Psi_{\MSK}(t) = e^{-j(\theta_o + s_d(t)\pi f_b t)}
     \label{eq:MSK},
\end{equation}
where $\theta_o$ is the final phase value from previous chip interval. In the next section, DPSK and MSK waveforms will be used to solve the constrained optimization problem in (\ref{eq:COP}).  

\section{Proposed Closed-form Receive Filter Design}
\label{Sec:Proposed}
To solve the constrained optimization problem defined in (\ref{eq:COP}), an iterative algorithm is proposed in \cite{DFRC_Rx_2010}. The proposed algorithm reduces the RSLs but does not achieve good coherency. Therefore, the algorithm is good to suppress range side-lobe levels but not good for Doppler shift measurement. However, the proposed solution in \cite{DFRC_Rx_2010} can achieve fully coherent RSLs for only two waveforms. This means, one can only transmit waveforms having only two bits of information. Similarly,  in \cite{shannonfilter2017} the same group of \cite{DFRC_Rx_2010}, for fixed system parameters using a continuous phase modulation (CPM) framework \cite{RC_DFRC_shannon_Waveforms1}, and based on match and mismatch filters proposed a cascaded filter design. Though, using these cascaded filters, one can enhance the coherency to reduce the RSMs but at the expense of fixed system parameters and higher sidelobe levels (that can reduce the sensing capability of the radar system). The motivation of our proposed work is to provide a generalized closed-form solution applicable for multiple waveforms, thus achieving the capacity to communicate any number of information bits and without sacrificing the performance of a DFRC system. 

To design our proposed receive filters, suppose there are $K$ waveforms as expressed in (\ref{eq:DPSK}) and (\ref{eq:MSK}). Considering $x_k(p)$ and $h_k(l)$ are $p$th and $l$th sampled values of $\Psi_k(t)$ and $h_k(t)$, respectively. Using sample values, the vector-form of convolution in (\ref{eq:COP}) can be written as 
\begin{eqnarray}
{\bf \Psi}_k{\bf h}_{k} = {\bf e}_{k},  
\label{eq:LinearModel}
\end{eqnarray}
where 
\begin{eqnarray}
{\bf \Psi}_{k} &=& 
\begin{bmatrix}
x_k(0)       & 0        & \cdots       &  0      \\
x_k(1)       & x_k(0)     & \ddots       &  \vdots \\
\vdots     & x_k(1)     & \ddots       &   0      \\
x_k(L-1)     & \vdots   & \ddots       & x_k(0)    \\
0          & x_k(L-1)   &              & x_k(1)    \\
\vdots     & \ddots   & \ddots       & \vdots  \\
0          & \cdots   &  0       & x_k(L-1)   
\end{bmatrix} \notag \\
{\bf h}_{k} &=&
\begin{bmatrix}
h_k(0) & h_k(1) & \cdots & h_k(L_{f}-1)\end{bmatrix}^T,~~~\mbox{and} \notag\\
{\bf e}_{k} &=&\begin{bmatrix}
0 & 0 & \cdots & 1 & \cdots & 0 & 0
\end{bmatrix}^T, \label{eq:COPVec}
\end{eqnarray}
where ${\Psi}_{k}\in {\cal C}^{(L+L_f-1)\times L_f}$. Similarly, the constraints in (\ref{eq:COP}) can also be written as 
\begin{align}
    {\bf \Psi}_1\mathbf{h}_{1}={\bf \Psi}_{2}\mathbf{h}_{2}=\cdots ={\bf \Psi}_{K}\mathbf{h}_{K}.
    \label{eq:constraint_lin}
\end{align}
Using (\ref{eq:COPVec}) and (\ref{eq:constraint_lin}), to design receive filters for $K$ waveforms, the objective function in (\ref{eq:COP}) can be defined as
\begin{eqnarray}
{\bf Xh = e} \label{eq:ObjK}\\
\mbox{and  }
\tilde{\bf X}{\bf h = 0} \label{eq:ConK},
\end{eqnarray}
where
\begin{eqnarray}
\mathbf{X} &=& \begin{bmatrix}
    {\bf \Psi}_{1} & \mathbf{0} & \cdots& \mathbf{0}\\
    \mathbf{0} &  {\bf \Psi}_{2} &  \cdots & \mathbf{0}\\
    \vdots& \vdots &\ddots & \vdots\\
    \mathbf{0} &\mathbf{0} & \cdots & {\bf \Psi}_{K}
\end{bmatrix}, \notag \\ 
\tilde{\bf X} &=& \begin{bmatrix}
{\bf \Psi}_{1} & -{\bf \Psi}_{2} &\mathbf{0} & \cdots& \mathbf{0} &\mathbf{0}\\
\mathbf{0} &  {\bf \Psi}_{2} &-{\bf \Psi}_{3}&  \cdots & \mathbf{0}&\mathbf{0}\\
\vdots& \vdots& \vdots &\ddots & \vdots& \vdots\\
\mathbf{0} &\mathbf{0}&\mathbf{0} & \cdots & {\bf \Psi}_{K-1}&-{\bf \Psi}_{K}
\end{bmatrix}, \notag \\
\mathbf{h} &=& \begin{bmatrix} \mathbf{h}_{1}^T & \mathbf{h}_{2}^T & \cdots \mathbf{h}_{K}^T
\end{bmatrix}^T, \notag\\
\mathbf{e} &=& \begin{bmatrix} \mathbf{e}_{1}^T & \mathbf{e}_{2}^T & \cdots \mathbf{e}_{K}^T\end{bmatrix}^T, \notag\
\label{eq:BlkMatrx}
\end{eqnarray}
where ${\bf X} \in {\cal C}^{(K(L+L_f-1))\times KL_f }$ and $\tilde{\bf X} \in {\cal C}^{(K-1)(L+L_f-1)\times KL_f}$. 

Using (\ref{eq:ObjK}) and (\ref{eq:ConK}), the constrained optimization problem to estimate ${\bf h}$ can be redefined as
\begin{eqnarray}
&&\underset{{\bf h}}{\text{minimize}}
\vectonorm{
     {{\bf X}{\bf h}-{\bf e}}
}^{2}\notag \\
&&\text{subject to} \notag \\
&& \qquad\qquad {\mathbf{\Tilde{X}}{\bf h}= {\bf 0}}. \label{eq:OptPr1}
\end{eqnarray}
The above constrained optimization problem can be solved using the Lagrangian multiplier, and the cost-function to be minimised can be written as 

\begin{eqnarray}
J =&  \vectonorm{
     {\mathbf{Xh-e}}
}^{2} + {\bf h}^H\tilde{\bf X}^H{\bm \lambda},  
\label{eq:CostFunction}
\end{eqnarray}
where vector ${\bm \lambda}$ is called Lagrangian multiplier. The minimization of (\ref{eq:CostFunction}) with respect to ${\bf h}$ yields
\begin{equation}
{\bf h} = \left({\bf X}^H{\bf X}\right)^{-1}\left[ {\bf X}^H{\bf e} - \tilde{\bf X}^H{\bm \lambda}\right]
\label{eq:h1}
\end{equation}
Substituting (\ref{eq:h1}) back into (\ref{eq:CostFunction}) yields 
\begin{align}
{\bm \lambda} &= {\bf D}^{-1}\tilde{\bf X}\left({\bf X}^H{\bf X}\right)^{-1}{\bf X}^H{\bf e},
\label{eq:lambda}
\end{align}
where ${\bf D} = \left( \tilde{\bf X}\left({\bf X}^H{\bf X}\right)^{-1}\tilde{\bf X}^H\right)$. Now substituting (\ref{eq:lambda}) into (\ref{eq:h1}) we get
\begin{equation}
  {\bf h} = \left({\bf X}^H{\bf X}\right)^{-1} {\bf X}^H{\bf e} - \left({\bf X}^H{\bf X}\right)^{-1}{\tilde{\bf X}^H}{\bf D}^{-1}{\tilde{\bf X}\left({\bf X}^H{\bf X}\right)^{-1}}{\bf X}^H{\bf e}\label{eq:filter_linear}.
\end{equation}
\\
{
{\bf Remark:} The derived estimation vector ${\bf h}$ in (\ref{eq:filter_linear}) contains two inverse matrices, i.e. $({\bf X}^H{\bf X})^{-1}$ and ${\bf D}^{-1} = \tilde{\bf X}\left({\bf X}^H{\bf X}\right)^{-1}\tilde{\bf X}^H$. 
To make the matrix $({\bf X}^H{\bf X})$ non-singular, the dimensions of ${\bf X}$ i.e, ${\cal C}^{(K(L+L_f-1))\times KL_f }$  must fulfill the bound $KL_f \le K(L+L_f-1)$. Similarly, for the matrix ${\bf D}$ to be full rank, the dimensions of $\tilde{\bf X} \in {\cal C}^{(K-1)(L+L_f-1)\times KL_f}$ must fulfill the bound $(K-1)(L+L_f-1)\le KL_f $ and this simply impose $L \le \frac{L_f}{K-1} +1$. Another way to express it for ill condition avoidance is $(K-1)(L+L_{f}-1)\leq KL_f\leq K(L+L_{f}-1)$.\\\\ 
}

In (\ref{eq:OptPr1}), instead of using tall linear convolution matrix, ${\bf \Psi}_k$, using square circular convolution matrix, ${\bf \Psi}_{ck}$, can be defined as 
\begin{eqnarray}
{\bf \Psi}_{ck} 
&=&\begin{bmatrix}
x_k(0)     & 0   &  \cdots    &  \cdots   & x_k(1)  \\
 \vdots    & \ddots  & \ddots  & \vdots & \vdots  \\
x_k(L-1)   &  & x_k(0)  & \ddots  &  x_k(L-1) \\
 \vdots    & \ddots & \vdots  &  \ddots  & {\bf 0}  \\
{\bf 0}  &   \cdots     & x_k(L-1)  & \cdots  & x_k(0)
\end{bmatrix}, \notag
\end{eqnarray}
ill matrix condition can be removed.  
The corresponding desired output response of the designed receive filter, ${\bf h}_{ck}$, can be written as 
\begin{eqnarray}
{\bf \Psi}_{ck}{\bf h}_{ck} = {\bf e}_{ck}. 
\label{eq:CircularModel}
\end{eqnarray}
In (\ref{eq:CircularModel}), ${\bf \Psi}_{ck} \in {\cal C}^{(L+L_f-1)\times (L+L_f-1)}$. Following similar strategy used to generate ${\bf X}$ and $\tilde{\bf X}$ with ${\bf \Psi}_k$s, we can generate ${\bf X}_c$ and $\tilde{\bf X}_c$ with ${\bf \Psi}_{ck}$ and can find corresponding ${\bf h}_c$ by solving the following constrained optimization problem,
\begin{eqnarray}
&& \underset{\mathbf{h}_c}{\text{minimize}}
  \vectonorm{
     {\mathbf{X_{\textit{c}}h_{\textit{c}}-e_{\textit{c}}}}
}^{2} \notag\\
&& \text{subject to} \notag\\
&&  \qquad\qquad {\mathbf{\Tilde{X}_{\textit{c}}h_{\textit{c}}}= {\bf 0}},
\label{eq:OptPrblm_zeroforcing_cir}
\end{eqnarray}
where the dimension of ${\bf X}_{\textit{c}} \in {\cal C}^{(K(L+L_f-1))\times (K(L+L_f-1))}$ and $\tilde{\bf X}_{\textit{c}} \in {\cal C}^{(K-1)(L+L_f-1)\times K(L+L_f-1)}$, while the dimensions of $\bf {h_{\textit{c}}}$ and $\bf{e_{\textit{c}}}$ are adjusted accordingly. Solving (\ref{eq:OptPrblm_zeroforcing_cir}) yields,
\begin{eqnarray}
  {\bf h_{\textit{c}}} &=& \left({\bf X}^H_{\textit{c}}{\bf X}_{\textit{c}}\right)^{-1} {\bf X}^H_{\textit{c}}{\bf e}_{\textit{c}} - \notag\\
  && \left({\bf X}^H_{\textit{c}}{\bf X}_{\textit{c}}\right)^{-1}{\tilde{\bf X}}^H_{\textit{c}}{\bf D}^{-1}_{\textit{c}}{\tilde{\bf X}_{\textit{c}}\left({\bf X}^H_{\textit{c}}{\bf X}_{\textit{c}}\right)^{-1}}{\bf X}^H_{\textit{c}}{\bf e}_{\textit{c}}\label{eq:filter_circular},
\end{eqnarray}
where ${\bf D}_c = \left({\tilde{\bf X}}_{c}\left({\bf X}^H_{c}{\bf X}_{c}\right)^{-1}\tilde{\bf X}^H_{c}\right)$. 

\begin{figure*}[ht]
    \centering
    \includegraphics[width=7in,keepaspectratio]{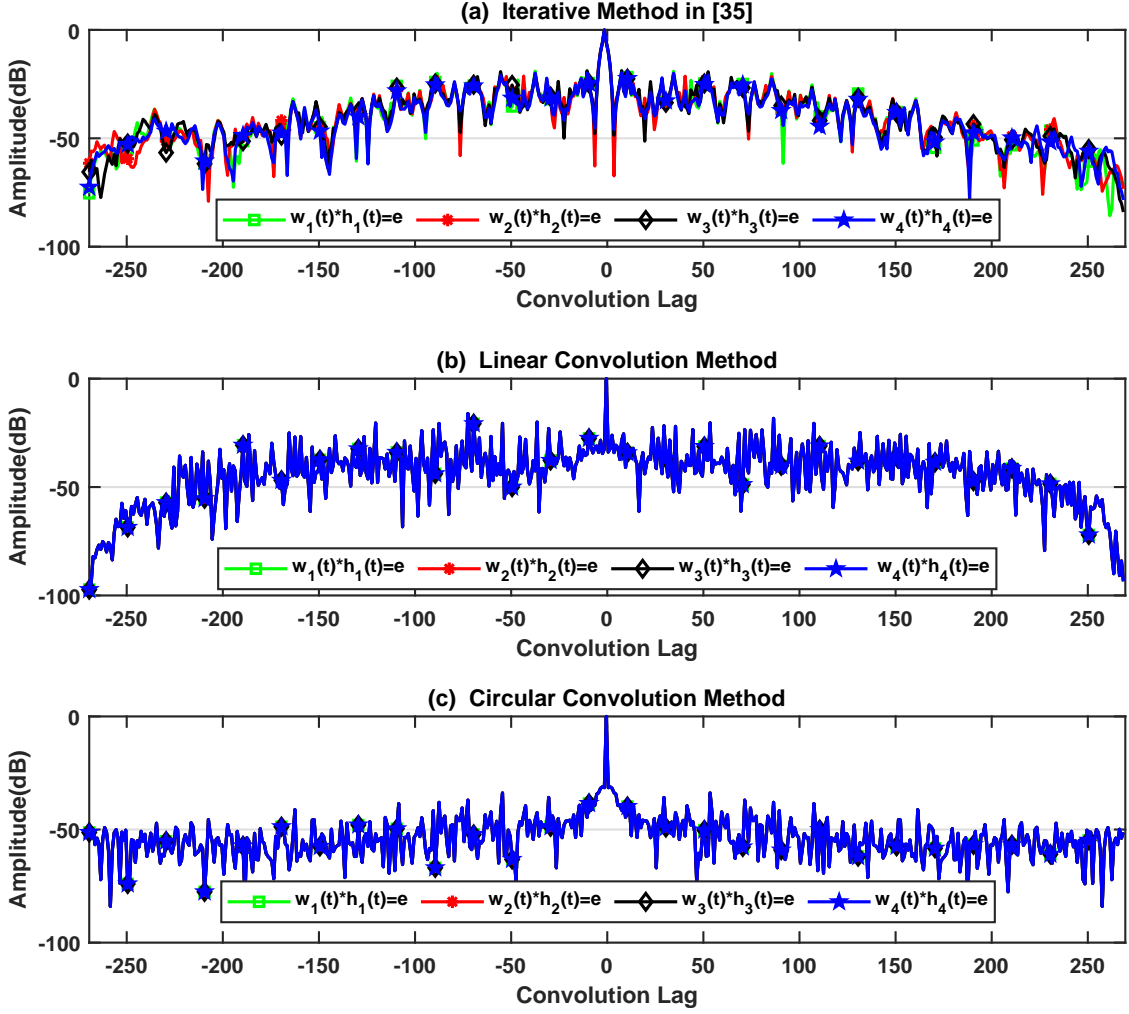}
    \caption{(a) Iterative receiver proposed in \cite{DFRC_Rx_2010}, RSL are not coherent, (b) first proposed receiver using linear convolution modelling, RSL are fully coherent and RSL levels are lover than \cite{DFRC_Rx_2010}, (c) second proposed receiver, RSL are fully coherent and RSL levels are lover than \cite{DFRC_Rx_2010}.}
    \label{fig:FiltersPerformance_[36]}
\end{figure*}
\section{Receive Signal Model}
\label{Sec:ClutterModel}
Receive signal can be model in two perspective, i.e., radar and communication. For communication, generally different frequencies are transmitted for up- and down-link channels. Therefore, for communication, an independent receiver can be modeled at the base-station. The design of receiver for radar is a challenging task, therefore, in the following discussion more focus will be on it.  
\subsection{Receive Signal Model for Radar}
Suppose a DFRC system, as shown in Fig.\ref{fig:KAUST}, transmitting $M$ waveforms separated by a PRI of $T_{\textit{PRI}}$.  
Consider, we have $P$ scaterers located at different ranges and angles. Some of the scatterers are stationery, some are  slow moving, while some of them are fast moving. We are interested in finding the fast moving ones. The stationary scatterers can be termed as clutter, while the slow moving and fast moving can be categorized as the targets. Consider a scatterer $i$ at the range $R_i$ from the transmitter. Since transmitter and receiver are co-located, the echo from the target will be received after time $\tau_i=\frac{2R_i}{c}$. 
When $m$th pulse is sent the range of the target will be $R_{i}\pm mv_{i}T_{\textit{PRI}}$ depending the target is moving toward or away from the base-station. Note that, it is assumed that the target movement is negligible during NCPI but there will be a phase shift after each symbol, it is valid assumption because EM wave moves with the speed of light \cite{richards2010doppler}. For such scenarios, the received signal due to the $m$th transmitted signal, defined in (\ref{eq:pasbnd1}), by ignoring noise can be written as
\begin{align}
r_m(t) =  \sum_{i=1}^{P}\beta_i \Psi_p\left(t-\frac{2(R_{i}-m{v_{i}}T_{\PRI})}{c}\right),  
\label{eq:ClutMod1}
\end{align}
where $m = 0,\ldots, M-1$ and $\beta_i$ is the reflection coefficient of $i$th scaterer. Using (\ref{eq:pasbnd1}), (\ref{eq:ClutMod1}) can be written as 
%
{
\begin{eqnarray}
    r_m(t) &=&  \sum_{i=1}^{P}\beta_i\bigg[ \tilde{s}_c\left(t-\tau_i-\frac{\tau_{c}}{2}\right)\cos\left(2\pi f_ct + \phi(t,R_i) \right) \notag \\
    &&\quad\quad~ -\tilde{s}_s\left(t-\tau_i\right)\sin \left(2\pi f_ct +\phi(t,R_i) \right)\bigg],
    \label{eq:receivedSignal}
\end{eqnarray}
}
%
where $\phi(t,R_i) = 2\pi f_{di} mT_{\PRI} - \frac{4\pi R_i}{\lambda}$ and ${ \tau_{i}\leq t\leq \tau_{i} +\tau}$ while $\tau_{i}=\frac{2(R_{i}-m{v_{i}}T_{PRI})}{c}$. Demodulating (\ref{eq:receivedSignal}), in-phase, $I(t)$, and quadrature-phase, $Q(t)$, components of the baseband received signal due to the clutter can be written as  
\begin{eqnarray*}
I(t) &=& \sum_{i=1}^{P}\beta_i\bigg[ s_{\mbox{\tiny{c}}}\left(t-\tau_{i}-\frac{\tau_{c}}{2}\right)\sin\left(2\pi f_{d}mT_{\textit{PRI}}-\frac{4\pi R_{i}}{\lambda} \right)\notag \\&&+s_{\mbox{\tiny{s}}}\left(t+\tau_{i}\right)\cos\left(2\pi f_{d}mT_{\textit{PRI}}-\frac{4\pi R_{i}}{\lambda} \right)\bigg].\\
    Q(t) &=& \sum_{i=1}^{P}\beta_i\bigg[ s_{\mbox{\tiny{c}}}\left(t-\tau_{i}-\frac{\tau_{c}}{2}\right)\cos\left(2\pi f_{d}mT_{\textit{PRI}}-\frac{4\pi R_{i}}{\lambda} \right)\notag \\&& -s_{\mbox{\tiny{s}}}\left(t-\tau_{i}\right) \sin\left(2\pi f_{d}mT_{\textit{PRI}}-\frac{4\pi R_{i}}{\lambda} \right)\bigg].
    \label{eq:Icomponent}
    \label{eq:Qcomponent}
\end{eqnarray*}
Now combining $I(t)$ and $Q(t)$, the baseband received signal due to the clutter can be written as
\begin{align}
y_{m}(t) = I(t) + jQ(t). \label{eq:receivedBaseBand}
\end{align}
Further solving (\ref{eq:receivedBaseBand}), the baseband received echo can be expressed as  
\begin{equation}
  y_{m}(t) = \sum_{i=1}^{P}\beta_i \Psi_\DPSK\left(t-\tau_{i}\right)
       e^{j(2\pi f_{di}mT_{\textit{PRI}}-\frac{4\pi R_{i}}{\lambda} )}.
    \label{eq:ComplexReceived1}
\end{equation}
In discrete vector form, $y_{m}(t)$ can be written as %
\begin{align}
      {\bf y}_m  = { \Psi}_{m} \left({\bm{\beta}}\odot {\mathbf{s}}(\phi) \odot  {\mathbf a}(f_d) \right) \label{eq:ReceivedVector},
\end{align}
where,
\begin{eqnarray}
{\bf y}_m &=& \begin{bmatrix}
y_m(0) & y_m(1) & \cdots & y_m(L+L_f-1)
\end{bmatrix}^T, \notag\\
{\bf \Psi}_{m} &=& 
\begin{bmatrix}
x_m(0)       & 0        & \cdots       &  0      \\
x_m(1)       & x_m(0)     & \ddots       &  \vdots \\
\vdots     & x_m(1)     & \ddots       &         \\
x_m(L-1)     & \vdots   & \ddots       & x_m(0)    \\
0          & x_m(L-1)   &              & x_m(1)    \\
\vdots     & \ddots   & \ddots       & \vdots  \\
0          & \cdots   & \cdots       & x_m(L-1)   
\end{bmatrix}, \notag \\
{\bm{\beta} } &=&
\begin{bmatrix}
\beta_{1} & \beta_{1} & \cdots & \beta_{P}\end{bmatrix}^T, \notag\\
{\bf s}(\phi) &=&\begin{bmatrix}
e^{-\frac{j4\pi R_1}{\lambda}} &e^{-\frac{j4\pi R_2}{\lambda}}  & \cdots & e^{-\frac{j4\pi R_{P}}{\lambda}}
\end{bmatrix}^T, \notag\\
{\bf a}(f_d) &=&\begin{bmatrix}
e^{j2\pi f_{d1}mT_{\textit{PRI}}}  & \cdots & e^{j2\pi f_{dP}mT_{\textit{PRI}}}
\end{bmatrix}^T.\notag
\end{eqnarray}
In (\ref{eq:ReceivedVector}), ${\bf \Psi}_m$ is the convolution matrix of $m$th waveform, ${\bm \beta}$ is the vector of reflection coefficients of $P$ scatterers, ${\bf s}(\phi)$ is the phase delay vector, and ${\bf a}(f_d)$ is the Doppler vector contains the Doppler shift of different scatterers. 
\subsection{Receive Signal Model for Communication}
The receive signal model for communication system at the base-station or user end is very straight forward. The baseband receive signal for the $m$th transmitted waveform can be written as
\begin{eqnarray}
c_m(t) = \sum_{l=0}^{L-1} g_le^{j2\pi f_d t} x_m(t) + v(t),
\end{eqnarray}
where $L$ is the number of paths, $g_l$ is the path gain of $l$th path, and $f_d$ is the Doppler shift due to the relative motion between the transmitter and receiver. The transmitted symbols and channel can be easily found using the conventional techniques \cite{Sajid_TC_2005}. 
\begin{figure*}
    \begin{centering}
    \includegraphics[width=1\textwidth]{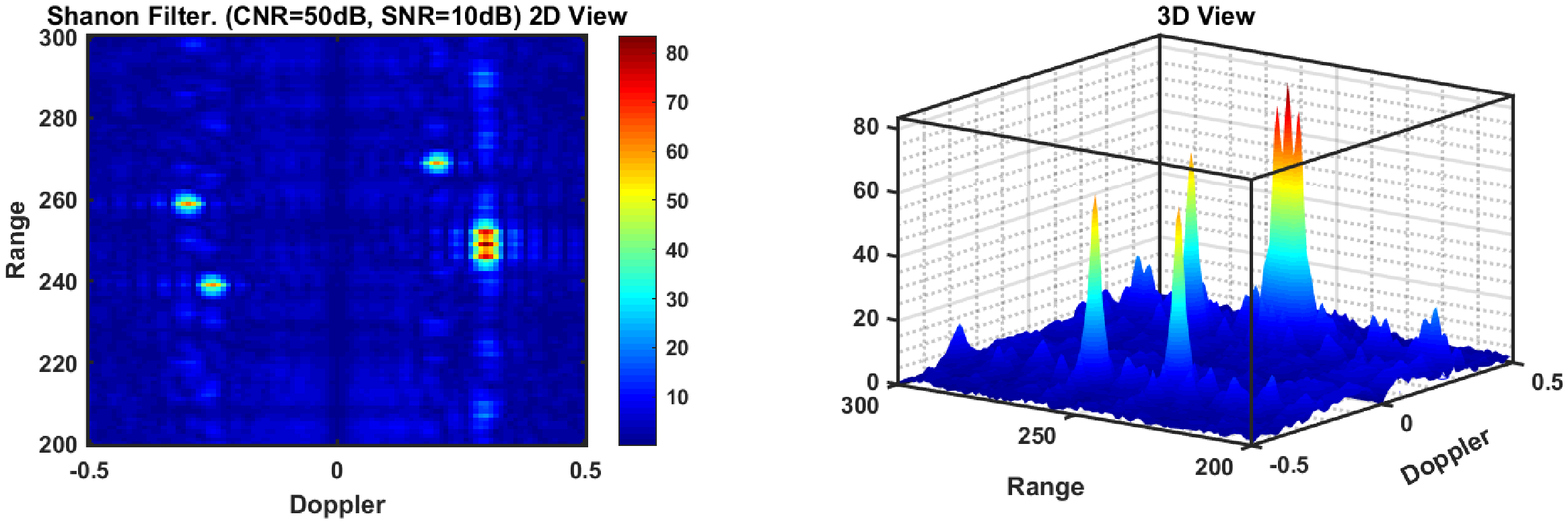}\\
    (a) \\  
    \includegraphics[width=1\textwidth]{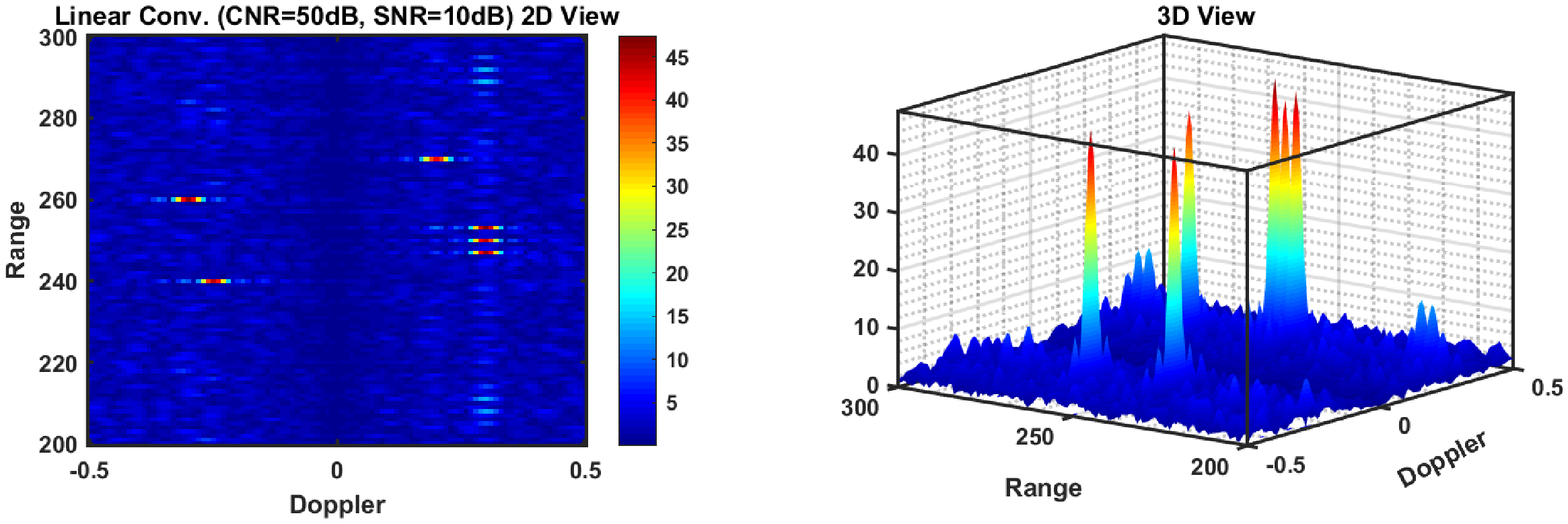}\\
    \vspace{1em}
    (b) \\
    \includegraphics[width=1\textwidth]{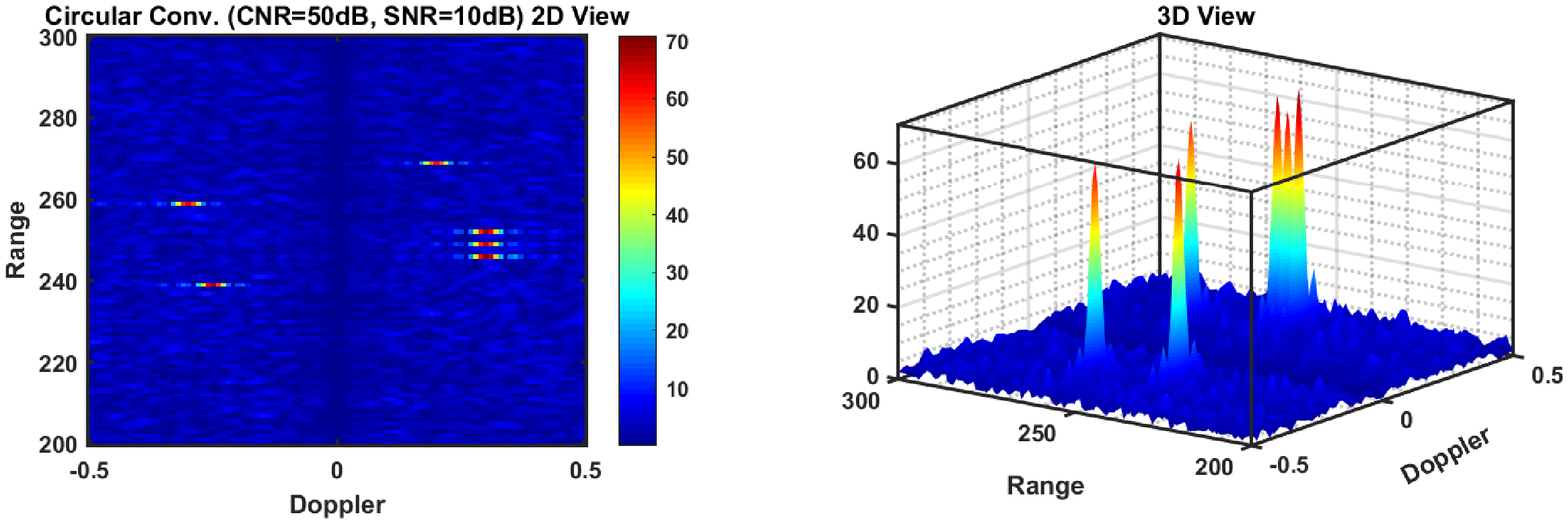}\\
    \vspace{1em}
    (c) \\ 
    \caption{Detection of six targets at CNR = $50$dB and SNR $10$dB. (a) Iterative receiver proposed in \cite{DFRC_Rx_2010}, closely spaced targets are difficult to distinguish, (b) 1st proposed receiver, all targets can be easily identified, (c) 2nd  proposed receiver, all targets can be easily identified.}\label{fig:myfig1}
    \end{centering}
\end{figure*}
\begin{figure*}
    \begin{centering}
    \includegraphics[width=\textwidth,height=2in]{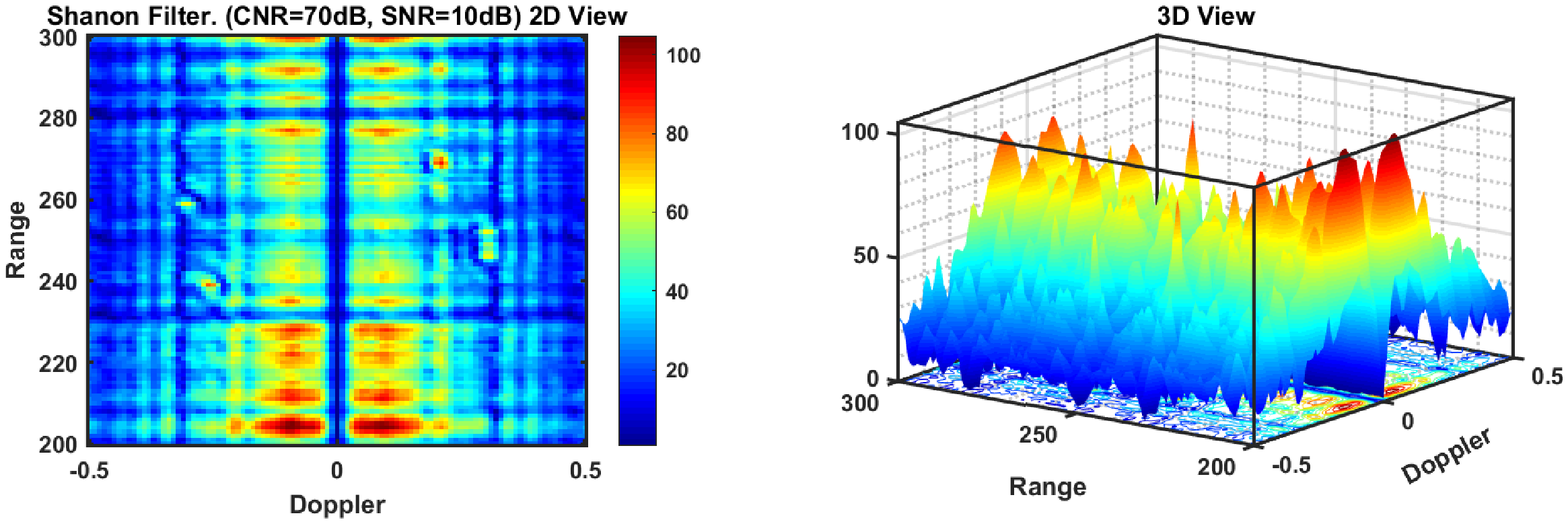}
    (a)  
    \includegraphics[width=\textwidth,height=2in]{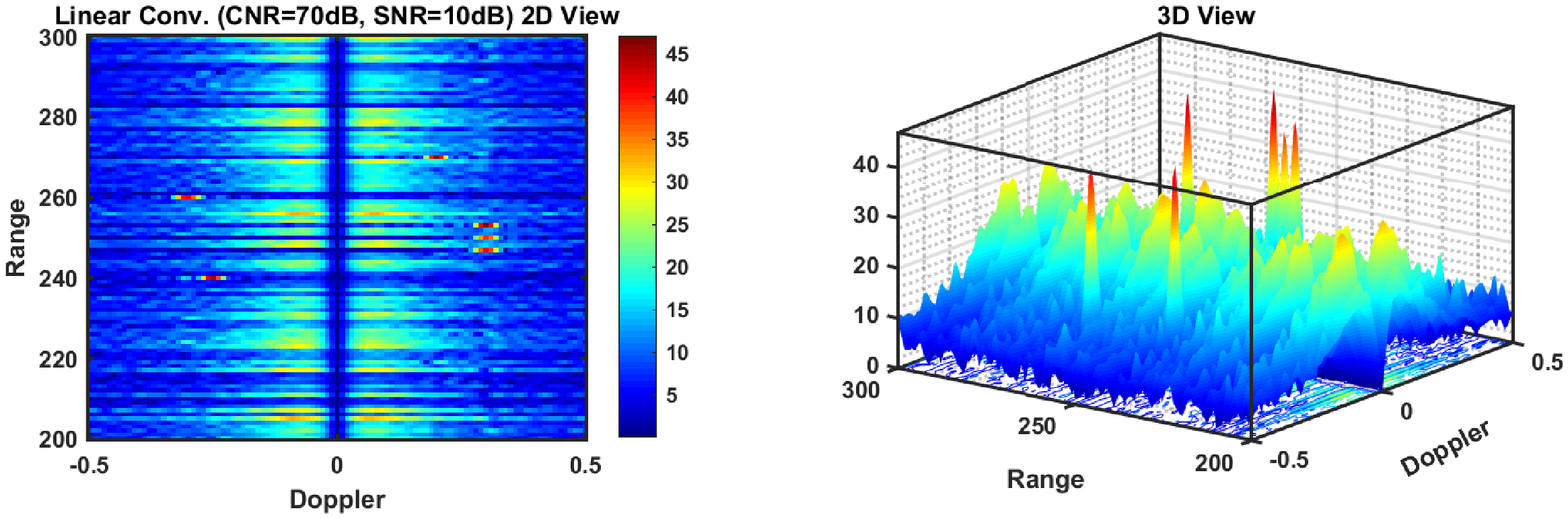}
    \vspace{1em}
    (b)
    \includegraphics[width=\textwidth,height=2in]{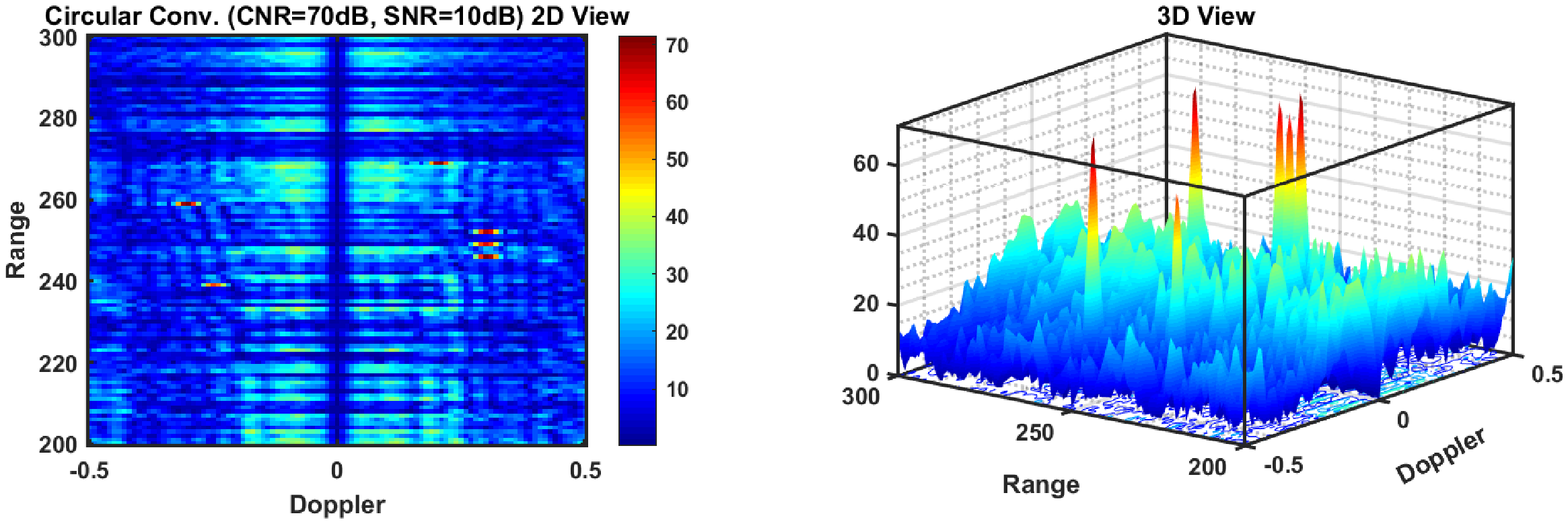}
    \vspace{1em}
    (c) 
    \caption{Detection of six targets at CNR = $70$dB and SNR $10$dB. (a) Iterative receiver proposed in \cite{DFRC_Rx_2010}, targets can not be identified, (b) 1st proposed receiver, all targets can be easily identified, (c) 2nd  proposed receiver, all targets can be easily identified.}\label{fig:myfig2}
    \end{centering}
\end{figure*}

\section{Simulation Results}
In this section, to assess the performance of our designed closed-form receive filters, a number of simulations are performed. For all simulation the number of waveforms $K=4$ and waveforms are DPSK. 

In the first simulation, we compare the RSL levels of our proposed receive filter with the iterative joint least-square (JLS) receive filter proposed in \cite{DFRC_Rx_2010}. Here, we consider the waveforms are DPSK and the number of waveforms is four. The simulation results are shown in Fig. \ref{fig:FiltersPerformance_[36]}. It can be observed from Fig. \ref{fig:FiltersPerformance_[36]}a, the responses of receive filters corresponding to different waveforms are not fully coherent and the computational complexity is high due to its iterative nature. Therefore, this receiver can mask week targets in the presence of a clutter, and the estimation of Doppler can be a challenging task. In Fig. \ref{fig:FiltersPerformance_[36]}b, the output of our designed receive filters using linear convolution is shown. As can be seen in the figure that the output of all filters is fully coherent and the RSL levels are comparatively lower than the iterative JLS receiver filter. Similarly, the output of our designed filters using circulant convolution technique is shown in Fig. \ref{fig:FiltersPerformance_[36]}c. It can be seen from \ref{fig:FiltersPerformance_[36]}c that the output of all receivers is fully coherent and RSLs are remarkably lower then both of the above techniques. 
%

In the second simulation, we compare the target detection performance of the proposed receivers with the iterative JLS receive filter proposed in \cite{DFRC_Rx_2010}. For this simulation,  $M=50$ waveforms are transmitted in each NCPI. Each waveform is randomly selected from 4 DPSK alphabet waveforms. The number of chips in each waveform is $N=30$, and each chip is of duration $\tau_c=1$ms. The total duration of each DPSK waveform is $T=30$ms. The pulse shaping sines and cosines in DPSK waveform have the baseband frequency $f_b=500$Hz, and their sampling rate is $3$kHz. There are 6 targets, 4 targets are moving towards the DFRC system while the remaining 2 targets are moving away. The normalised Doppler frequency of first 3 targets is $0.3$Hz and fourth target is $0.25$Hz. The overall range of our DFRC system is divided into $450$ range gates, and fast moving scatterers (targets) are placed around the central range gates. For example, 3 targets with Doppler shift of $0.3$Hz are located in range cells\footnote{The smallest range increment the radar is capable of detecting.}225, 228 and 221. The other three targets with Doppler frequencies of $-0.3$Hz, $-0.25$Hz and $0.2$Hz are located in range cells of 235, 215, and 245, respectively. 
Putting received samples corresponding to different waveforms in different columns can form a matrix. Samples in a particular row of this matrix are called slow time samples and they represent echoes from a target in the same range-bin. Applying fast-Fourier-transform (FFT) on the matrix in the row dimension can find the Doppler shift of moving targets in different range-bins. The corresponding FFT results are shown in Fig. \ref{fig:myfig1} and \ref{fig:myfig2}. Note that only $100$ central range gates are shown in the simulation results for better visualization. Echos from slowly moving and stationary objects are considered as clutter returns. In simulations, the clutter is modeled as a white Gaussian noise process and have the normalized Doppler range of $-0.1$Hz to $0.1$Hz for stationary and slowly moving scatterers. For simulation two scenarios are considered, in the first scenario the signal-to-noise ratio (SNR) is $10$dB and clutter-to-noise ratio (CNR) is $50$dB, while in the second scenario SNR\;$=10$dB and CNR\;$ = 70$\;dB. The corresponding performances are shown in Fig.\;\ref{fig:myfig1} and  Fig.\;\ref{fig:myfig2}. 
%
For the first scenario, Fig.\;\ref{fig:myfig1}a shows that with JLS iterative receive filter, the three closely located targets in rang-gates 225, 228, and 221 each of Doppler shift $0.3$Hz are poorly resolvable. The other three widely separated targets are resolvable, but due to the non-coherence response of the receive filter, targets are heavily masked by nearby clutters. In contrast, Fig.\;\ref{fig:myfig1}b shows the detection performance of our proposed linear-convolution-based receiver. As can be seen in the figure, the closely located targets are resolved efficiently due to the fully coherent response of our proposed receiver. Similarly, the performance of our proposed circular-convolution-based receiver is shown in Fig. \ref{fig:myfig2}, which is even better than the linear-convolution-based receiver (please compare the amplitudes of peaks). 

Similarly, Fig.\ref{fig:myfig2} shows the detection performance of all receivers for the second scenario. It can be seen in Fig.\ref{fig:myfig2}a, the JLS iterative receiver completely fails to detect the targets in the presence of nearby strong clutters. On the other hand, Fig. \ref{fig:myfig2}b shows the linear-convolution-based receivers have successfully resolved and unmasked the targets buried in heavy clutter. Similarly, Fig.\;\ref{fig:myfig2}c shows that the circular-convolution-based receivers, as in the first scenario, have significantly better performance compared to JLS and linear-convolution-based receivers. 

\begin{figure}
    \centering
    \includegraphics[width=3.53in,height=2.75in]{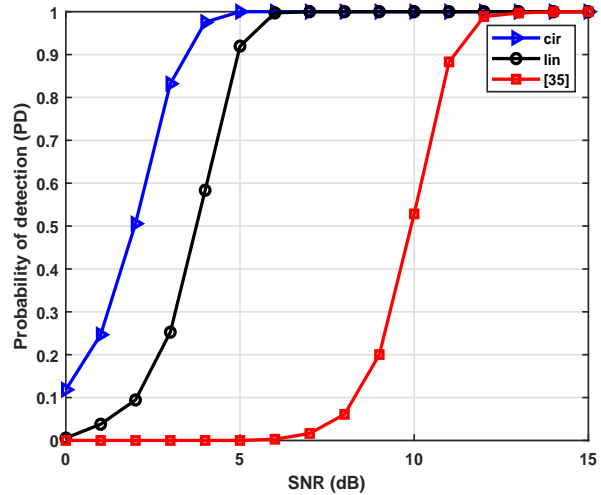}
    \caption{Probability of detection comparison of proposed schemes with the JLS iterative receiver proposed in \cite{DFRC_Rx_2010}.}
    \label{fig:my_label}
\end{figure}

In the last simulation of radar, the probability of detection of proposed receivers is compared with the JLS iterative receiver. Fig.\;\ref{fig:my_label} shows that the performance of proposed linear- and circular-convolution-based receivers is much better than the JLS iterative receiver.    

\begin{figure}
    \centering
    \includegraphics[width=3.53in,height=2.75in]{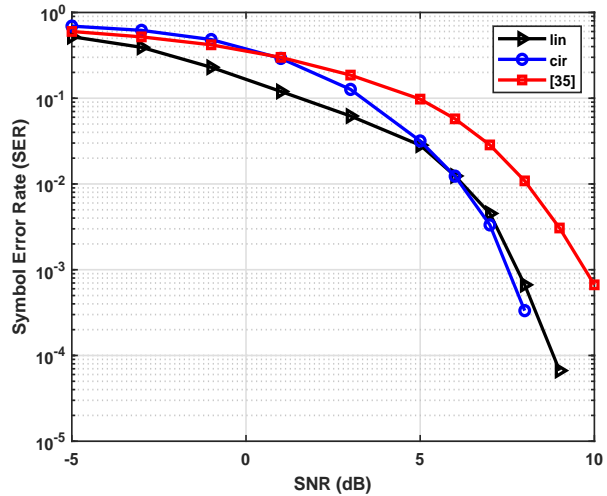}
    \caption{SER comparison of our proposed schemes with the JLS iterative receiver proposed in  \cite{DFRC_Rx_2010}.}
    \label{fig:my_SER}
\end{figure}

Finally, we perform a simulation to find the symbol-error-rate (SER) of the communication. For simplicity, we consider a line-of-sight communication (single-path channel). However, simulations can be easily extended for the multipath channel. Here, the received waveform (symbol) is passed through all receive filters. For detection of the symbol, the receiver that yields maximum output amplitude, its corresponding waveform is considered as the detected symbol. 
Since we have used only four waveforms, each can carry only two bits. Fig. \ref{fig:my_SER} compares the SER rate, where it can be seen that the proposed liner convolution based receiver performs the best for SNR lower than 5dB. While for higher SNR i.e., $SNR\geq5$dB, circular-convolution based receiver performs the best. This performance behavior of our proposed schemes is because at higher SNR they provide better cross-correlation among received waveform and receivers. At a typical SER of $10^{-2}$, both proposed liner-convolution and circular-convolution based receivers yield a 2dB gain in SNR compared to the iterative JLS receiver \cite{DFRC_Rx_2010}. This gain is because, during optimization, JLS iterative cost-function does not have to fulfill any constraint as in our proposed schemes. As a consequence, compared to our proposed design, the target detection and SER of the JLS iterative scheme are poor.             
\section{Conclusions}{\label{Sec:Simulations}}
In this paper, to reduce range side-lobes and achieve full coherency, two novel closed-form algorithms based on linear- and circular-convolution are proposed. The computationally complexity of the proposed algorithms is lower than the proposed algorithms available in the literature and achieves full coherency for any number of waveforms. As far as we know no one has achieved full coherence for more than two waveforms. Simulation results demonstrate the superiority of our proposed algorithms over the others for target detection and symbol-error-rate (SER). In the future work, the focus will be to improve the cross-correlation among different received waveform and their corresponding receivers.          
\bibliographystyle{ieeetr}
\bibliography{JournalRadar}

\end{document}